\documentstyle[preprint,aps,psfig,emlines2]{revtex}

\begin{document}

\title{How many Raman-active vibrations do exist  \\
in carbon nanotubes?}

\author{T.~Yu.~Astakhova, and G.~A.~Vinogradov}

\address{Institute of Biochemical Physics RAS, ul.Kosygina~4,
Moscow~GSP-1~119991, Russia \\
E-mail: astakhova@deom.chph.ras.ru;\,\, gvin@deom.chph.ras.ru }

\maketitle

\begin{abstract}
In this paper we present the results on Raman-active modes
calculation in achiral $(n,n)$ and $(n,0)$ carbon nanotubes (CNTs).
vibrational spectra are derived as the eigenvalues of corresponding
dynamical matrix at the $\Gamma$-point of Brillouin zone.  Diagonal
components and $\alpha _{xy} \, (=\alpha _{yx})$ are the only
non-zero components of polarization tensor. Selection rules for
Raman-active modes are determined by direct estimations of matrix
elements responsible for the intensities of corresponding vibrational
transitions. Few $E_{2g}$ modes can be non-Raman-active because
of their ``fine'' vibrational structure. We claim that numbers of
Raman-active modes for achiral CNTs are: 5 (\,$(n,n)$ and $(n,0)$,
$n$-even\,); 6 (\,$(n,n)$, $n$-odd\,); 8 (\,$(n,0)$, $n$-odd\,).

\end{abstract}

%%%%%%%%%%%%%%%%%%%%%%%%%%%%%%%%%%%%%%%%%%%%%%%%
%section 1
\section{Introduction}
\label{intro}

Carbon nanotubes (CNTs) are highly organized quasi-one-dimensional
structures consisting of rolled-up graphene sheets. The variations in
their diameters and helicities result in differences in electronic
and vibrational properties\cite{book_1}, \cite{book_2}, with Raman
spectroscopy providing a powerful tool for their characterization.
Currently, a few problems still exist in the calculation of the full
list of Raman-active vibrations and their symmetry assignments.  For
some time it was believed that there were 15 or 16 Raman vibrations
for achiral $(n,0)$ and $(n,n)$ CNTs depending on whether $n$ is odd
or even\cite{Jishi_1}.  The group-theoretical analysis predicts 7
intensive Raman-active vibrations in the range up to
1600~$cm^{-1}$\cite{Saito_3}. Decomposition of experimentally
observed spectra into individual components was made following this
predictions\cite{Rao_1}, \cite{Journet}.

Recently the total number of Raman-active vibrations was declared to
be half of the previous estimates, namely 8 for both armchair and zigzag
CNTs\cite{Alon}.  The corresponding calculations are based on the rod
(or line) group theory\cite{Damnjanovic}, worked out earlier for the
stereoregular polymers\cite{Milosevic}.

In the present work we perform a detailed analysis of the vibrational
spectra of infinite CNTs with a view to address the unresolved
issues.  Our approach differs from the conventional approach, where
the point and spatial symmetry groups are assigned first, followed by
the full spectral analysis. Calculations are performed using
Brenner's empirical potential for hydrocarbons\cite{Brenner}.
Infinite CNTs are simulated by supercells with periodic boundary
conditions. The full set of normal vibrational modes is obtained by
the diagonalization of the dynamical matrix. The mode symmetries are
assigned according to commonly accepted rules\cite{Flurry},
\cite{Elyashevich}, \cite{Wilson}. Selection rules for Raman-active
modes are found by an estimation of matrix elements responsible for
the intensity of the corresponding mode. The estimation is based on
the direct comparison of the mode symmetry with the symmetry of
polarization tensor components in polar coordinates. This approach
does not require the preliminary complete group-theoretical analysis
of CNTs. Nevertheless, the necessary symmetry assignment of vibrational
modes is made.

In this paper we present results for armchair $(n,n)$ and zigzag
$(n,0)$ CNTs with $n = 9 - 12$. The detailed analysis of the
vibrational modes demonstrates that there can be modes with identical
symmetries that give different contributions to Raman spectra. It
follows from the difference in the ``fine'' structure of these modes,
and depends on whether vibrations of neighboring atoms are
``in-phase'' or ``out-of-phase''.

The manuscript is organized as follows: In the second section we
present the details of the construction and analysis of the dynamical
matrix.  Sec.~ \ref{symmetry} sketches the symmetry properties of
achiral nanotubes.  Sec.~\ref{raman} provides a brief description of
the general rules for the estimation of the Raman-active vibrations.
In Sec.~\ref{results} we present results for achiral tubes $(n,n)$
and $(n,0)$ where the vibrational spectrum of a $(10,10)$ nanotube is
analyzed in more details as an example. Sec.~\ref{concl} provides a
summary of our results.

%%%%%%%%%%%%%%%%%%%%%%%%%%%%%%%%%%%%%%%%%%%%%%%%%%%%%%%%%%%%%%%%%%%%%%%
%section 2
\section{Calculation of vibrational spectrum using the dynamical
matrix.}
\label{dynamic}

The vibrational spectrum can be derived as the solution of the matrix
equation:

\begin{equation}
\label{m_eq}
{\bf D}\,X = M\,\omega ^2 {\bf B}\,X ,
\end{equation}

\noindent
where ${\bf D}$ is the dynamical matrix and ${\bf B}$ is the matrix of
coefficients of quadratic form of kinetic energy  $E_{kin} =
\frac{1}{2}\sum b_{ij} \, v_i \, v_j $, ($v_i$, $v_j$ are generalized
velocities of atoms $i$ and $j$).  $\{ \omega _k \}$ and $\{X_k\}$
are, respectively, the eigenvalues and eigenvectors fully describing
the vibrational spectrum.  $M$ is the mass of carbon atom. Matrix
elements ${\bf D}_{ij}$ are given by:

\begin{equation}
\label{D_matrix}
{\bf D}_{ij}^{\alpha \beta}=
{\displaystyle \frac{\partial ^2 E}{\partial \alpha _i \, \partial
\beta _j}},
\end{equation}

\noindent
where $E$ is the potential energy of the system; $\alpha _i,
\beta _j$ are generalized coordinates of atoms $i$ and $j$. In
Cartesian coordinates ${\bf B}$ in (\ref{m_eq}) is the unit matrix
and vibrational spectrum is obtained by diagonalization of matrix
${\bf D}$. Corresponding eigenvectors contain all information on
symmetry and relative amplitudes of atomic displacements from the
equilibrium.  The Brenner's potential\cite{Brenner} is used for
modeling nanotube structure and vibrational properties. Analytical
form of this potential allows to get second derivatives analytically.
Cylindrical coordinates greatly facilitate the construction of the
dynamical matrix: it is sufficient to calculate the matrix elements
for two neighboring atoms and then the dynamical matrix is
reconstructed by symmetry.  The aspect ratio of CNTs is very large
and it allows to consider the CNTs as infinite. We use a supercell
with periodic boundary conditions to model the infinite tube.

If a supercell consists of $N$ atoms, the rank of the dynamical
matrix is $3\,N \times 3\,N$ and Eqn.~(\ref{m_eq}) has $3\,N$
solutions giving the full set of normal vibrations.  We consider
modes only in the center of the Brillouin zone ($\Gamma $-point with
$k_z = 0$). Modes with $k_z \neq 0$ give contribution to
Raman-spectra for finite CNTs \cite{Saito_1}, \cite{Richter}, and
will be ommited in the present consideration.  As a result, the rank
of the dynamical matrix decreases to $3\,N'\times 3\,N'$, where
$3\,N'$ is the number of vibrations with $k_z = 0$.

The symmetry assignments are standard (see, {\it e.g.},
\cite{Flurry}, \cite{Elyashevich}, \cite{Wilson}). For
one-dimensional representations letters $A$ and $B$ stand for
symmetric and antisymmetric modes with respect to $S^1_{2n}$ symmetry
operations; subscripts `1' and `2' - denote symmetric and
antisymmetric modes with respect to the rotation about two-fold axes
perpendicular to the principal axis. The letters $`g'$ and $`u'$, --
symmetric and antisymmetric modes with respect to the inversion. In
two-dimensional representations $E_k$, the subscript $k$ stands for
modes which transform to itself by the $S^k_{2n}$ symmetry operation.
An advantage for the choice of $S_{2n}$ as the principal axis over
the $C_n$ axis for CNTs is given in Sec.~\ref{results}. The results
are summarized in Tables~I, II and will be discussed below.

There are $A$ and $B$ modes with ``double'' symmetries with respect
to the rotation about $C_2$ axes perpendicular to the principal axis
$S_{2n}$.  An example of such (Raman-inactive) mode in $(n,n)$ CNT
with even $n$ is given in Fig.~1.  The mode shown is the $B_g$
tangential mode where all atoms in every layer\footnote{All atoms
with equal z-coordinates form a layer.} have equal displacements.
There are three classes of $C_2$ rotational axes perpendicular to
the tube axis ($C'_2$, $C''_2$, $C'''_2$) for even $n$. (More
detailed analysis of CNT symmetry elements will be given in the next
Section).  One can see that mode $B_g$ is antisymmetric with respect
to $C'_2$ and $C''_2$ axes and is symmetric with respect to $C'''_2$
axis. This is the reason why subscripts `1' or `2' are omitted for
some $A$ and $B$ modes in Table~I.

Following these rules, the symmetry analysis shows that the vibrational
modes of $(n,n)$ and $(n,0)$ CNTs with even and odd $n$ have
following symmetries:

\begin{equation}
\label{list}
\begin{array}{ll}
\Gamma _{(n,n)}^{n-even} = & 2\,A_{1g} +
2\,A_{2g} + 2\,A_u +1\,B_{1u} + 1\,B_{2u} + 4\,B_g + \\
& 2\,E_{1g} + 4\,E_{1u} + 4\,E_{2g} + 2\,E_{2u} + \ldots  +
  2\,E_{(n-1)g} + 4\,E_{(n-1)u} \\
\Gamma _{(n,n)}^{n-odd} =  & 2\,A_{1g} + 2\,A_{2g} + 2\,A_g +
                     1\,B_{1u} + 1\,B_{2u} + 4\,B_u +
   6\,E_{1u} + 6\,E_{2g} + \ldots  +
  6\,E_{(n-1)g} \\
\Gamma _{(n,0)}^{n-even} = & 2\,A_{1g} + 1\,A_{2g} + 3\,A_u +
                    1\,B_{1u} + 2\,B_{2u} + 3\,B_g + \\
 & 3\,E_{1g} + 3\,E_{1u} + 3\,E_{2g} + 3\,E_{2u} + \ldots  +
     3\,E_{(n-1)g} + 3\,E_{(n-1)u}\\
\Gamma _{(n,0)}^{n-odd} = & 2\,A_{1g} + 1\,A_{2g} + 3\,A_g +
                    1\,B_{1u} + 2\,B_{2u} + 3\,B_u +
  6\,E_{1u} + 6\,E_{2g} + \ldots  +  6\,E_{(n-1)g}

\end{array}
\end{equation}

\noindent
Four zero-frequency modes are not shown in (\ref{list}). The
rotation about $z$-axis has $A_{2g}$ symmetry (see Fig.~2a); the
translation along the $z$-axis has $B_{2u}$ symmetry \footnote{$B$
symmetry is assigned because this mode is antisymmetric with respect
to $S_{2n}^1$ operation which is a rotation by an angle $2 \pi / 2n$
and reflection in the horizontal plane  passing through centers of
the slanted bonds. If the $C_{n}$ axis is taken as the principal axis
then this mode has $A_{2u}$ symmetry since it is symmetric with
respect to the $C_n^1$ rotation (Fig.~2b)}; translations in
$(x,y)$ plane are doubly-degenerated and have $E_{1u}$ symmetry
(Fig.~2c). Symmetries of zero-frequency modes differ from symmetries
given in \cite{book_2} because we use $S_{2n}$ as the principal axis.

Before proceeding with the analysis of the Raman-activity of
calculated vibrational modes we briefly recall the main symmetry
elements in CNTs.

%%%%%%%%%%%%%%%%%%%%%%%%%%%%%%%%%%%%%%%%%%%%%%%%%%%%%%%%%%%%%%%%%%%

\section{Symmetry elements in CNTs.}
\label{symmetry}

An analysis of the vibrational modes and comparison with polarization
tensor components requires a knowledge of necessary symmetry elements
of CNTs. Note that the symmetry analysis is independent on the form
of interatomic potential; this choice determines only frequencies but
not symmetries.

The full group-theoretical analysis of CNTs is beyond the scope of
this work \footnote{One can find the group-theoretical analysis in
\cite{Alon}, and nonsimmorphic rod-group is:  ${G[n]=L_{T_z} \times
\left[D_{nh}|_{z=0} \otimes \left( D_{nd}|_{z=T_z/4} \ominus C_{nv}
\right) \oplus C_{nv} \times S_{2n} \right] }$.}.  A list of allowed
point symmetry elements for $(n,n)$ CNTs with even and odd $n$ is
given in Table~III.  Symmetry elements for $D_{nh}$ and $D_{nd}$
point groups are also shown for comparison.

Fig.~3 illustrates symmetry operations for $(n,n)$ CNTs with even and
odd $n$ ($n=6$ in Fig.~3a, and $n=5$ in Fig.~3b). Two different forms
of unit cells are used, since different symmetry elements are
clarified in different representations. Two top panels show chains of
hexagons chosen as unit cells. Top panel in column (b) (odd $n$)
shows a two-fold rotation axes, $C'_2$, passing through centers of a
hexagon and an opposite horizontal bond. This axis splits into two
classes of two-fold rotation axes for even $n$, as shown in the top
panel of column (a). One rotation axis, $C'_2$, passes through
centers of opposite hexagons, while the other, $C''_2$, passes
through centers of opposite horizontal bonds.  The $(n,n)$ CNT with
even $n$ is thus associated with two classes of reflection planes:
$\sigma ' _v$ and $\sigma '' _v$ passing through $C'_2$ and $C''_2$
axes, respectively.  By contrast, there is only one class of
reflection planes, $\sigma ' _v$ associated with a CNT with odd $n$.
Moreover, while the point $i$ in the top panel for even $n$ is an
inversion center, the corresponding point $O'$ in the right panel is
not an inversion center. Two top panels also show that both tubes
have horizontal reflection planes $\sigma _h$ passing through centers
of hexagons.

In order to illustrate other symmetry elements we show the armchair
fragments as the unit cell for both tubes in middle panels in
Fig.~3.  One can see that there are additional two-fold rotation
axes, $C'''_2$, passing through centers of opposite slanted bonds.
Additionally, both tubes have $2\, n$-fold roto-reflection axes,
$S_{2n}$.  The inversion center in the case of odd $n$ (right panel)
is $i$ and inversion is absent for even $n$.

The bottom panels in Fig.~3 are the side view of both tubes with
corresponding symmetry elements.

To summarize: there are symmetry elements, $C_n$, $S_{2n}$,
$\sigma _v$, $C'''_2$, as well as two types of fixed points, $i$
(inversion) and $O$, which are common to CNTs with both even and odd
$n$.  Symmetry elements $C'_2$, $C''_2$, $\sigma ' _v$,
$\sigma '' _v$ are inherent to CNTs with even $n$, while elements
$C'_2$ and $\sigma ' _v$ are characteristic to odd $n$.

This analysis also demonstrates that all symmetry elements of both
$D_{nd}$ and $D_{nh}$ groups are permitted in CNTs (see Table~III and
Fig.~3).  One can say that CNTs `combine' symmetry properties of
$D_{nh}$ and $D_{nd}$ point groups. In particular, CNT has three
classes of two-fold axes for even $n$:  ($C'_2$, $C''_2$, $C'''_2$).
These axes are the combination of two classes for $D_{nh}$ symmetry
group ($C'_2$, $C''_2$ axes) and one class of $C_2$ axes for $D_{nd}$
($C'''_2$ axes).

In the case of zigzag CNTs an analysis is similar. Thus,
with a knowledge of CNTs symmetry elements and the calculated
vibrational spectra we can identify Raman-active modes in CNTs.

%%%%%%%%%%%%%%%%%%%%%%%%%%%%%%%%%%%%%%%%%%%%%%%%%%%%%%%%%%%%%%%%%

\section{How to reveal Raman-active modes?}
\label{raman}

In this Section we recall main rules for determining Raman-active
modes.  Let the transition between states $E_i$ and $E_j$ is
associated with wave functions $\psi_i$ and $\psi_j$.  Then the
probability for the $\psi _i \to \psi _j$ process is proportional to
the square of the transition moment $M_{ij}$, where

\begin{equation}
\label{moment}
M_{ij} = \int \psi_{j}^{*} M \psi_i \, dV.
\end{equation}

\noindent
In Raman process, $M$ is the dipole moment induced by the external
field $\vec E$. The induced dipole moment is
$\vec P = \alpha \vec E$, where $\alpha $ is a polarization tensor.
Thus,  the selection rules in Raman spectra are determined by the
values of integrals

\begin{equation}
\label{probability}
\int \psi_{j}^{*} \alpha_{\xi \, \xi '} \psi_i \, dV,
\end{equation}

\noindent
where $\alpha _{\xi \, \xi '}$ is a component (or a linear
combination of components) of the polarization tensor
($\xi, \, \xi ' = x,y,z$).  The transition $i \to j$ is allowed if
(\ref{probability}) is nonzero for at least one component
$\alpha_{\xi \, \xi '}$. The integration is performed over the full
set of electronic and nuclear coordinates.  Direct calculation of
(\ref{probability}) is very complicated problem. So, the usual
practice is to consider the electronic, vibrational and rotational
degrees of freedom separately.  Thus the wave function can be written
as the product $\psi = \psi^{el}\,\psi ^{vib} \, \psi ^{rot}$. As a
result only the following integrals should be estimated for Raman
transitions:

\begin{equation}
\label{probability_v}
\int \psi_{j}^{vib} \alpha_{\xi \, \xi '} \psi_i^{vib} \, dV.
\end{equation}

vibrational wave function $\psi ^{vib}$ is real and is represented by
an eigenvector of the dynamical matrix. Since we consider the
$\Gamma$-point of Brillouin zone, where the dependence on the $z$
coordinate is absent, the space integration in (\ref{probability_v})
is replaced by the area integral

\begin{equation}
\label{probability_s}
\int \psi_{j}^{vib} \alpha_{\xi \, \xi '} \psi_i^{vib} \,dx\,dy.
\end{equation}

It is convenient to return back to the polar coordinates
$(\xi , \, \xi ' = r, \, \phi )$. In the simplest case the
ground-state vibrational mode $\psi _i$ is fully symmetric and is a
constant in polar coordinates. Moreover, the vibrational wave
function is the discrete function (eigenvector of the dynamical
matrix). Then (\ref{probability_s}) transforms to the sum:

\begin{equation}
\label{probability_sum}
\sum _{m=1}^{2n}\psi_j^m \alpha_{\xi \, \xi '} ,
\end{equation}

\noindent
where the summation is over atoms $m, \,\,\, m=1,2, \ldots 2\,n$ of only
one layer, and products of polarization tensor components with $m$-th
elements $\psi _j^m$ of eigenvector of $j$-th normal mode are
calculated for angles $\phi ^m $, corresponding to polar coordinate
of atom $m$. This approach is very convenient as both polarization
tensor components and eigenvector elements of vibrational modes have
very simple representation in polar coordinates for CNTs.

As is demonstrated in Appendix, only $\alpha _{xx}$, $\alpha _{yy}$,
$\alpha _{xx} \pm \alpha _{yy}$ and $\alpha _{xy}$ components of
polarization tensor differ from zero for CNTs and are shown in Fig.~4.

Many one-dimensional modes in (\ref{list}) can be excluded from
candidates for Raman vibrations. (i) All modes with subscript `u' are
eliminated as all components of polarization tensor are symmetric
with respect to this operation; (ii) All $B$ modes are excluded as
antisymmetric with respect to $S_{2n}$ operation. (iii) $A_{2g}$
modes are also non-Raman-active as they are asymmetric with respect
to $C_2$ operations. Hence, only two fully symmetrical $A_{1g}$ modes
are Raman-active.

Doubly-degenerated $E_{kg}$ modes in polar coordinates behave as
$\sin(k\phi + \psi)$, where $\psi $ -- some phase. Their schematic
view is given in Fig.~5.  The signs assignment is arbitrary. For
instance, the clockwise direction for tangential displacements and
radial displacements out of center have `+' sign.

Comparison of Fig.~4 and Fig.~5 results in the fact that only
$E_{2g}$ mode can be Raman-active.  Really, the shapes of components
of polarization tensor coincide only with the shape of $E_{2g}$ mode,
and signs of their lobes alternate in the same manner.  Actually, it
is necessary to estimate expression (\ref{probability_s}), which  can
be reduced to

\begin{equation}
\label{probability_E_kg}
\int _0^{2\,\pi} \sin(2\,\phi +\theta)\,\sin(k\,\phi + \psi)\,d\phi,
\end{equation}

\noindent
and is different from zero only if $k=2$. ($\sin(2\,\phi +\theta$) is
the general representation for $\alpha _{xy}$ and $\alpha _{xx -
\alpha _{yy}}$ components of polarization tensor, and phases $\psi $
-- for vibrational modes and $\theta $ -- for polarization tensor are
related as their principal axes coincide.). It means that only
$E_{2g}$ modes can be Raman-active. The conclusion that only $A_{1g}$
and $E_{2g}$ modes are Raman-active is valid for all types of achiral
CNTs, both armchair and zigzag with arbitrary $n$ (see Table~II).

%%%%%%%%%%%%%%%%%%%%%%%%%%%%%%%%%%%%%%%%%%%%%%%%%%%%%%%%%%%%%%%%

\section{Raman-active modes in achiral CNTs}
\label{results}

Two $A_{1g}$ modes for all types of CNTs are Raman-active as they are
fully symmetrical ``breathing'' and tangential modes (see Fig.~6).
Now we will demonstrate that one of $E_{2g}$ modes is not
Raman-active on the example of $(10,10)$ CNT. All $E_{2g}$ modes are
shown in Fig.~7. One can see the difference between mode (c) and
three other modes, this deference being in `signs' of relative
displacements of neighboring atoms. These displacements are
predominantly ``in-phase'' for (a), (b) and (d) modes and are
``out-of-phase'' for (c) mode.  Hence, these ``phase'' relations of
different modes give different contributions to
(\ref{probability_sum}): this sum is not equal to zero for modes (a),
(b), (c) and is zero (or very close to zero) for mode (c). It means
that mode (c) is not Raman-active, and Raman spectra of $(10,10)$ CNT
have 5 Raman-active vibrations (see Table~II). These conclusion is
valid for all $(n,n)$ CNTs with even $n$.

Situation with $E_{2g}$ vibrations for $(n,n)$ CNTs with odd $n$ is
analogous. Among six $E_{2g}$ modes there are two ``out-of-phase''
modes which are not Raman-active, and total number of Raman-active
vibrations in this type of CNTs is 6.

There are no ``out-of-phase'' modes in $(n,0)$ CNTs, and the total
number of Raman-active vibrations is the sum of $A_{1g}$ and $E_{2g}$
mode types: 5 Raman-active modes for $(n,0)$ CNTs if $n$ is odd, and
8 Raman-active modes if $n$ is even.

The reason why armchair CNTs differ from zigzag CNTs consists in the
fact that the latter have one atom in the primitive cell, and the
former -- two atoms. Two atoms in the primitive cell for armchair
CNTs allow intrinsic vibrational ``fine'' structure. Qualitatively
it can be demonstrated as follows.

We consider $E_{kg}$ modes, and the armchair ring is chosen as the
unit cell. This unit cell consists of $4\,n$ atoms from two layers
(see Fig.~8a, where $(6,6)$ CNT is shown, as an example).  Let us
open this fragment into a planar structure (Fig.~8b) and further
project into a linear chain (Fig.~8c).  There are two atoms in the
primitive cell (two adjacent atoms in a layer) since $S_{2n}$ axis
was chosen as a principal axis.  So every filled circle in Fig.~8c
represents one primitive cell of two atoms.  This gives a chain with
cyclic boundary conditions consisting of $n_{c} = 2\,n$ units, and
its solution is convenient to write in the form:

\begin{equation}
\label{linear_chain}
u(k,l) = a \, \cos (2\,\pi \, k \, l /n_c),
\end{equation}

\noindent
where $u(k,l)$ is a deviation of the unit $l$ for $k$-th type of
normal vibration ($k = 0,1, \ldots n_c-1$), and $l$ numerates the
units in the chain ($l=1,2, \ldots n_c$).

So, the full vibrational spectrum of $(n,n)$ CNT in the center of the
Brillouin zone can be approximated by a solution of the
one-dimensional chain with cyclic boundary conditions and some
effective potential.

The solution (\ref{linear_chain}) gives all one- and two-dimensional
modes. Really, if $k=0$ then all units have equal deviations
$u(0,\,l) = a$ of the same sign, which corresponds to $A$ modes. If
$k=n_c/2$ then all units have equal absolute values of deviations,
but the signs of relative deviations alternate $u(n_c/2,\,l) =
(-1)^l\,a$, which corresponds to $B$ modes.

Modes $E_{kg}$ are formed by pairs of solutions (\ref{linear_chain})
with `conjugated' values of $k$: $k$ and $n_c-k$.  Really,

$a \,\cos (2\,\pi\,(n_c-k)\,l /n_{c }) =
 a \,\cos (2\,\pi\,k\,l /n_{c })$,

\noindent
and it gives $(n-1)$ $E_{kg}$ modes for $(n,n)$ CNTs in agreement
with (\ref{list}).

The solution (\ref{linear_chain}) for $k=2$ is shown in Fig.~8c
and corresponds to $E_{2g}$ mode of the nanotube. This mode can be
represented schematically in polar coordinates as shown in Fig.~8d
and is equivalent to mode $E_{2g}$ in Fig.~5.

%%%%%%%%%%%%%%%%%%%%%%%%%%%%%%%%%%%%%%%%%%%%%%%%%%%%%%%%%%%%%%%%%%%

\section{Conclusions}
\label{concl}

The main goal of the present communication is the calculation of
numbers of Raman-active modes in achiral CNTs $(n,n)$ and $(n,0)$
($n$ = 9--12). vibrational modes are calculated in $\Gamma$-point of
Brillouin zone, and Raman-active modes are identified by an
estimation of matrix elements responsible for the intensity of
corresponding transitions, and no preliminary full group-theoretical
analysis of CNTs is necessary in this case.  For achiral CNTs only
three diagonal and $\alpha _{xy} = \alpha _{yx}$ components of
polarization tensor are not equal to zero.

We have shown that there are one or two $E_{2g}$ modes for $(n,n)$
CNTs which are not Raman-active. It follows from the fact that the
primitive cell of these CNTs has two atoms, and vibrational modes
have ``in-phase'' and ``out-of-phase'' types of vibrations, the
latter being non-Raman-active.  The primitive cell of $(n,0)$ CNTs
consists of one atom and these CNTs has no analogous complications.

The calculated numbers of Raman-active modes are five or six for
$(n,n)$ CNTs for even and odd $n$, correspondingly. For $(n,0)$ CNTs
there are five Raman-active modes for even $n$ and eight Raman-active
modes for odd $n$.

Our primary goal was to formulate the selection rules for
Raman-active modes and less attention was paid to calculations of
mode frequencies. More precise values of mode energies can be
obtained using more sophisticated potentials, {it e.g.} Menon's
tight-binding scheme \cite{Menon_1}, \cite{Menon_2}.

\vspace{1cm}

The authors thank the RFBR (Project 02-02-16205) and INTAS (Project
00-237) for the partial financial support. T.A. is also indebted to
RFBR (Project 00-15-97334). Stimulating discussions with M.Menon,
E.Richter, L.A.Gribov, and B.N.Mavrin are gratefully acknowledged.

%%%%%%%%%%%%%%%%%%%%%%%%%%%%%%%%%%%%%%%%%%%%%%%%%%%%%%%%%%%%%%

\appendix
\section {Polarization tensor components for CNTs}

1) We consider non-resonance Raman transitions. For simplicity
achiral CNT are represented as homogeneous hollow cylinder of
sufficiently large radius.

2) Let external electric field is applied along the $x$-axis
perpendicular to the vertical nanotube axis $z$. This field can not
induce polarization along $z$-axis as the ``top'' and the ``bottom''
of CNT are  equivalent. By symmetry the same is valid for
$y$-direction of applied field.  Hence, $\alpha _{xz} = \alpha _{zx}
= \alpha _{yz}= \alpha _{zy} = 0$.

3) Now we define the angular dependence of components $\alpha _{xx}$
and $\alpha  _{xy}$ in polar coordinates. External electric field
is aligned along the $x$-axis and we consider the $(xy)$
section of CNT (see Fig.~9) . Polarization vector at point
$A, \,\, (\phi =0)$ has only normal component: $\vec {P_n} =
\alpha _n \, \vec E$ , and $\alpha _n$ is the normal polarizability.
There is only tangential component at point $B$ $(\phi = \pi/2)$ and
$\vec P_{\tau}=\alpha _{\tau}\,\vec E$, where $\alpha _{\tau}$ --
tangential polarizability.  Obviously, $\alpha _{\tau} > \alpha _n$
as polarizability in the plain is larger compared to perpendicular
direction.

4) Consider now point $C$ at an angle $\phi $ to $x$-axis. Electric
field at this point has normal ($E_n=E\,\cos \phi$) and tangential
($E_{\tau}=E\,\sin \phi$) components. Normal and tangential components
of polarization vector at this point are:

\begin{equation}
\label{appendix_1}
\left \{
\begin{array}{ll}
P'_n       = & \alpha _n\,E\,\cos \phi             \\
P'_{\tau}  = & \alpha _{\tau}\,E\,\sin \phi
\end{array}
\right. ,
\end{equation}

\noindent
and $x$- and $y$-components of polarization vector $\vec{P'}$ at
point $C$ are:

\begin{equation}
\label{appendix_2}
\left \{
\begin{array}{ll}
P'_x = & P'_n\,\cos\phi + P'_{\tau}\,\sin\phi   \\
P'_y = & P'_n\,\sin\phi - P'_{\tau}\,\cos\phi
\end{array}
\right.
\end{equation}

Substituting in system (\ref{appendix_2}) values for $P'_n$ and
$P'_{\tau}$ from (\ref{appendix_1}) one gets:

\begin{equation}
\label{appendix_3}
\left\{
\begin{array}{ll}
P'_x = &(\alpha_n\,\cos^2\phi + \alpha _{\tau}\,\sin ^2\phi)\,E \\
P'_y = &(\alpha_n\,\sin \phi\,\cos\phi - \alpha _{\tau}\,\sin
\phi\,\cos \phi)\,E
\end{array}
\right.
\end{equation}

\noindent
And it follows that

\begin{equation}
\label{appendix_4}
\left\{
\begin{array}{ll}
\alpha _{xx}= &\alpha _n \, \cos^2\phi + \alpha _{\tau}\,\sin^2\phi \\
\alpha _{xy}=&-\frac{1}{2}\,(\alpha _{\tau}-\alpha _n)\,\sin2\,\phi
\end{array}
\right.
\end{equation}

One can get expressions for other non-zero components of polarization
tensor in polar coordinates:

\begin{equation}
\label{appendix_5}
\left\{
\begin{array}{rll}
\alpha _{yy}              & = & \alpha _{\tau} \, \cos^2\phi +
                                \alpha _n\,\sin^2\phi           \\
\alpha _{xx}+\alpha _{yy} & = & \alpha _{\tau}+\alpha _n        \\
\alpha _{xx}-\alpha _{yy} & = & (\alpha _n-\alpha _{\tau})\,
                                \cos 2\,\phi
\end{array}
\right.
\end{equation}

Schematic view of all non-zero components of polarization tensor is
shown in Fig.~4.

%%%%%%%%%%%%%%%%%%%%%%%%%%%%%%%%%%%%%%%%%%%%%%%%%%%%%%%%%%%%%%%%

%%%%%%%%%%%%%%%%%%%%%%%%%%%%%%%%%%%%%%%%%%%%%%%%%%%%%%%%%%%%%%

\newpage

\baselineskip 12pt

\begin{table}
\caption{ The vibrational spectrum of $(10,10)$ CNT calculated using
the Brenner potential. The mode frequency $\omega $ (in cm${}^{-1}$)
and its symmetry $S$ are given in the first and second columns. The
polarization tensor elements $\alpha_ {\xi \xi '}$ are also shown.
Raman-active vibrations are underlined. }

\begin{tabular}{|rll|rll|rll|}
\hline
$\omega$ & $S$ & $\alpha _{\xi \xi '}$ & $\omega$ & $S$
& $\alpha _{\xi \xi '}$ &
$\omega$ & $S$ & $\alpha _{\xi \xi '}$ \\
\hline
  {\underline {12}} & $E_{2g}$ & ($\alpha _{xx} - \alpha_{yy}$,
$\alpha_{xy}$) &

      523  & $E_{1g}$ &                                               &
     1415  & $E_{9u}$ &                                               \\
       34  & $E_{3u}$ &                                               &
      530  & $A_{u}$  &                                               &
     1464  & $E_{8g}$ &                                               \\
       65  & $E_{4g}$ &                                               &
      558  & $B_{g}$  &                                               &
     1515  & $E_{7u}$ &                                               \\
       81  & $E_{9g}$ &                                               &
      571  & $E_{9u}$ &                                               &
     1561  & $E_{6g}$ &                                               \\
      102  & $E_{5u}$ &                                               &
      596  & $E_{4g}$ &                                               &
     1600  & $E_{5u}$ &                                               \\
      144  & $E_{6g}$ &                                               &
      605  & $E_{8g}$ &                                               &
     1632  & $E_{4g}$ &                                               \\
 {\underline {150}} & $A_{1g}$ & $\alpha _{xx} + \alpha_{yy}$,
$\alpha_{zz}$   &

      648  & $E_{7u}$ &                                               &
     1656  & $E_{3u}$ &                                               \\
      160  & $E_{8u}$ &                                               &
      692  & $E_{6g}$ &                                               &
     1662  & $A_{u}$  &                                               \\
      189  & $E_{7u}$ &                                               &
      715  & $E_{5u}$ &                                               &
     1663  & $E_{1g}$ & ($\alpha_{xz}$, $\alpha_{yz}$)                \\
      211  & $E_{1u}$ &                                               &
      746  & $E_{5u}$ &                                               &
     1665  & $E_{2u}$ &                                               \\
      231  & $E_{8g}$ &                                               &
      774  & $E_{4g}$ &                                               &
     1669  & $E_{3g}$ &                                              \\
      236  & $E_{7g}$ &                                               &
      804  & $E_{3u}$ &                                               &
{\underline {1673}} & $E_{2g}$ & ($\alpha _{xx} - \alpha_{yy}$,
$\alpha_{xy}$) \\

      265  & $E_{9u}$ &                                               &
      826  & $E_{2g}$ &                                               &
     1673  & $E_{4u}$ &                                               \\
      279  & $A_{g }$ &                                               &
      840  & $E_{1u}$ &                                               &
     1677  & $E_{5g}$ &                                               \\
      307  & $E_{6u}$ &                                               &
      845  & $A_{2g}$ &                                               &
     1682  & $E_{6u}$ &                                               \\
 {\underline {331}} & $E_{2g}$ & ($\alpha _{xx} - \alpha_{yy}$,
$\alpha_{xy}$) &

      849  & $E_{6g}$ &                                               &
     1683  & $E_{1u}$ &                                               \\
      370  & $E_{5g}$ &                                               &
      961  & $E_{7u}$ &                                               &
     1685  & $E_{7g}$ &                                               \\
      425  & $E_{4u}$ &                                               &
     1060  & $E_{8g}$ &                                               &
{\underline {1686}} & $A_{1g}$ & $\alpha _{xx} + \alpha_{yy}$,
$\alpha_{zz}$   \\

      464  & $E_{3u}$ &                                               &
     1138  & $E_{9u}$ &                                               &
     1688  & $E_{8u}$ &                                               \\
      470  & $E_{3g}$ &                                               &
     1172  & $B_{g}$  &                                               &
     1690  & $E_{9g}$ &                                               \\
      503  & $E_{2u}$ &                                               &
     1391  & $B_{g}$  &                                               &
     1691  & $A_{1u}$ &                                               \\
\hline
\end{tabular}
\end{table}

\newpage
\begin{table}
\caption{ Raman-active modes of $(n,n)$ and $(n,0)$ CNT for $n=9-12$
calculated using the Brenner's potential.}

\begin{tabular}{|l|l|}
CNT & Frequencies (cm${}^{-1}$) and symmetries of Raman-active modes\\
\hline
(9,9)   & 15 ($E_{2g}$), 167 ($A_{1g}$), 366 ($E_{2g}$), 1665 ($E_{2g}$),
          1667 ($E_{2g}$), 1684 ($A_{1g}$) \\
(10,10) & 12 ($E_{2g}$), 150 ($A_{1g}$), 331 ($E_{2g}$), 1673 ($E_{2g}$),
          1686 ($A_{1g}$) \\
(11,11) & 10 ($E_{2g}$), 137 ($A_{1g}$), 302 ($E_{2g}$), 1665 ($E_{2g}$),
          1677 ($E_{2g}$), 1688 ($A_{1g}$) \\
(12,12) & 9 ($E_{2g}$), 125 ($A_{1g}$), 278 ($E_{2g}$), 1680 ($E_{2g}$),
          1689 ($A_{1g}$) \\
(9,0)   & 47 ($E_{2g}$), 286 ($A_{1g}$), 362 ($E_{2g}$), 591 ($E_{2g}$),
          741 ($E_{2g}$), 1144 ($E_{2g}$), 1672 ($E_{2g}$),
          1677 ($A_{1g}$)\\
(10,0)  & 38 ($E_{2g}$), 258 ($A_{1g}$), 542 ($E_{2g}$),
          1676 ($E_{2g}$), 1680 ($A_{1g}$)\\
(11,0)  & 31 ($E_{2g}$), 235 ($A_{1g}$), 348 ($E_{2g}$), 499 ($E_{2g}$),
          695 ($E_{2g}$), 1147 ($E_{2g}$), 1679 ($E_{2g}$),
          1683 ($A_{1g}$)\\
(12,0)  & 26 ($E_{2g}$), 216 ($A_{1g}$), 462 ($E_{2g}$),
          1681 ($E_{2g}$), 1685 ($A_{1g}$)\\
\end{tabular}
\end{table}

\newpage
\begin{table}

\label{sym_op}
\caption{
Symmetry operations for $D_{nh}$ and $D_{nd}$
point groups and symmetry operation for $(n,n)$ CNTs.
Here $E$ is the unit operation; $C_n^m$ is the rotation about the
z-axis by angle $2 \pi m /n$; $C'_2, C''_2,C'''_2$ are rotations about
$C'_2, C''_2, C'''_2$ axes (see Fig.~3) by an
angle $\pi$; $\sigma _h$ is the reflection in the plane
perpendicular to the  main symmetry axis; $\sigma _v$ is reflection
in the plane including the main symmetry and one of $C_2$
axes; $\sigma _d$ is reflection in the plane which
includes the principal axis and bisects the angle between two two-fold
axes; $S_n^m$ is the rotation about the principal axis by angle $2\pi
m/n$ and reflection in the plane $\sigma _h$: $S_n^m =
C_n^m \otimes \sigma _h$; $i$ - is the inversion operation.}

\begin{tabular}{|c|cccccc|}
\hline
& $D_{nh}$ & $D_{nh}$ & $D_{nd}$ & $D_{nd}$ & CNT     & CNT\\
&          &          &          &          & $(n,n)$ & $(n,n)$    \\
& even $n$ & odd $n$  & even $n$ & odd $n$  &  even $n$ & odd $n$  \\
\hline
$E$                             & +     & +   & +   & +   & +  & +\\
$C_n^m, m = 1,\ldots n-1$       & +     & +   & +   & +   & +  & +\\
$S_n^m, m = 1,\ldots n-1$       & +     & +   & -   & -   & +  & +\\
$S_{2n}^m, m = 1, 3\ldots 2n-1$ & -     & -   & +   & +   & +  & +\\
$\sigma_v$                      & +     & +   & -   & -   & +  & +\\
$\sigma_h$                      & +     & +   & -   & -   & +  & +\\
$\sigma_d$                      & +     & -   & +   & +   & +  & +\\
$C'_2$                          & -     & -   & +   & +   & +  & +\\
$C''_2$                         & +     & +   & -   & -   & +  & -\\
$C'''_2$                        & +     & -   & -   & -   & +  & +\\
$i$                             & +     & -   & -   & +   & +  & +\\
\hline
\end{tabular}

\end{table}

%%%%%%%%%%% -- FIGURE CAPTIONS --- %%%%%%%%%%%%%%%%

\newpage

\baselineskip 16pt

\centerline {FIGURE CAPTIONS}

\vspace{1.5cm}

\noindent
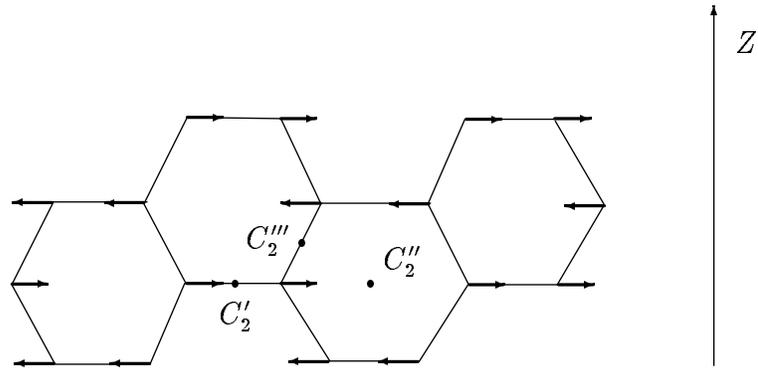
Fig.~1. An example of $B_g$ mode ($\omega$ = 1172 cm${}^{-1}$) with
``double'' symmetry is shown as a side view of an armchair CNT with
even $n$. The mode is antisymmetric with respect to rotation about
$C'_2$ and $C''_2$ axes and is symmetric with respect to rotation
about $C'''_2$ axis. Axes are perpendicular to the principal axis
$z$.  $C'_2$, $C''_2$ and $C'''_2$ axis pass through centers of
opposite horizontal bonds, centers of opposite hexagons, and centers
of opposite slanted bonds, respectively.

\vspace{0.5cm}

\noindent
Fig.~2. Zero-frequency modes: (a) $A_{2g}$ -- rotation about the
principal axis; (b) $B_{2u}$ -- translation along the principal axis;
(c) $E_{1u}$ -- translations along $x$ and $y$ axes perpendicular to
the principal axis $z$.

\vspace{0.5cm}

\noindent
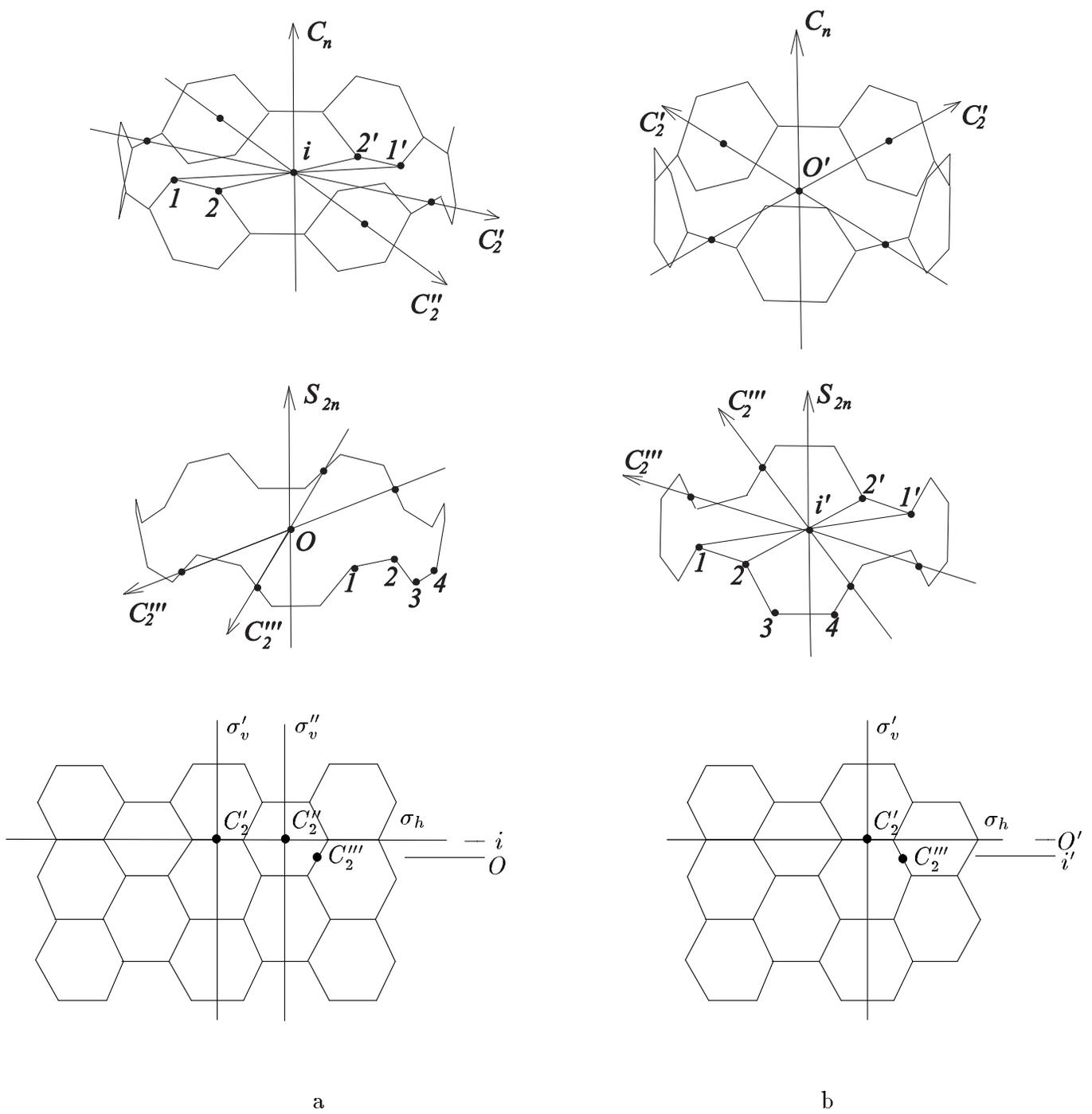
Fig.~3. Unit cells of $(6,6)$ (left column) and $(5,5)$ (right
column) CNTs with their symmetry elements. Top panels show unit cell
chosen as a chain of hexagons; middle panels present armchair unit
cells. The side view of tube fragments is given in the bottom panels.

\vspace{0.5cm}

\noindent
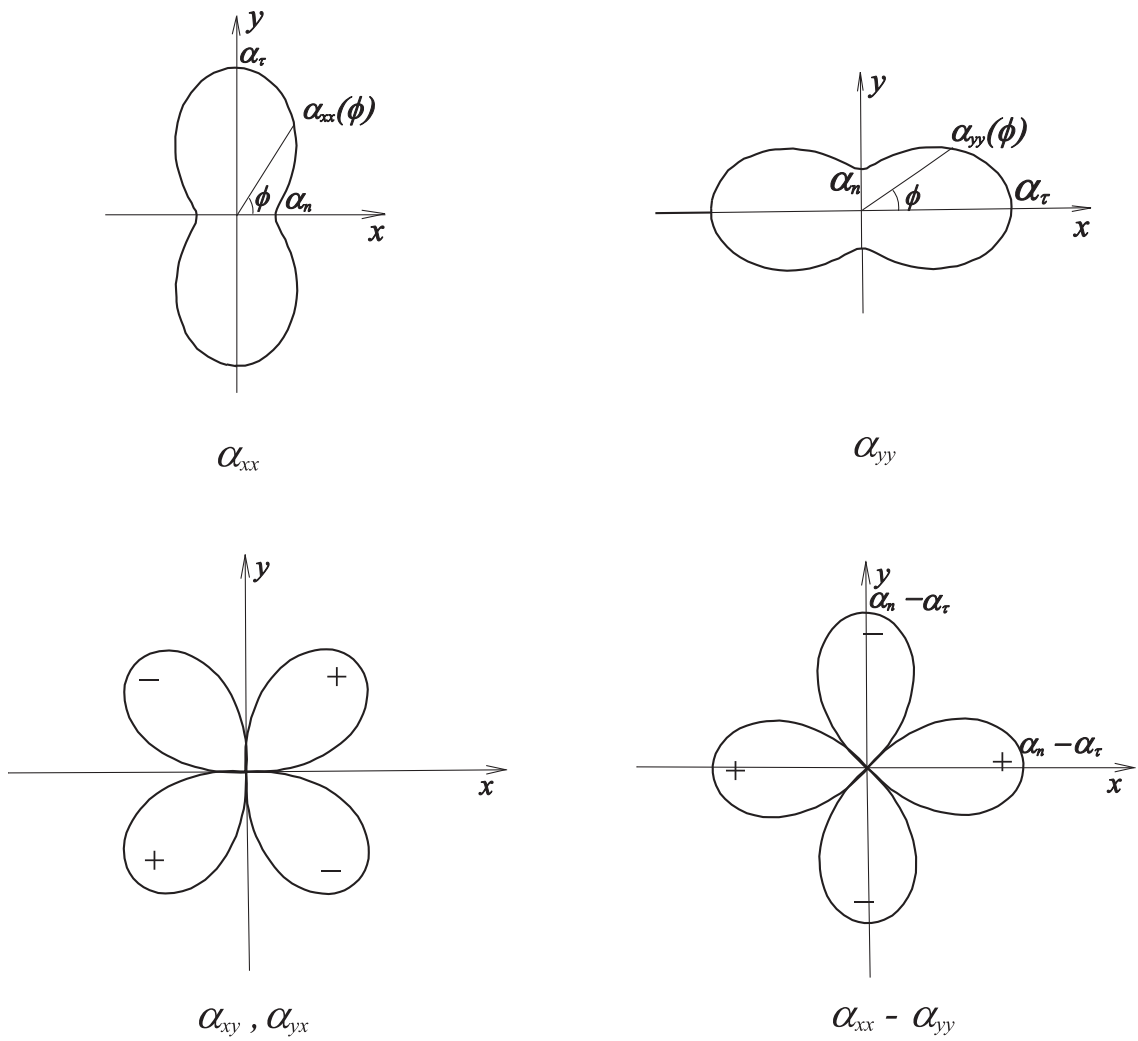
Fig.~4. Schematic representation of non-zero polarization matrix
elements in $(x,y)$ section of CNTs:
$\alpha_{xx} = a_n\,\cos^2\phi + a_{\tau}\,\sin^2\phi$,
$\alpha_{yy} = a_{\tau}\,cos^2\phi + a_n\,\sin^2\phi$,
$\alpha_{xy} = \alpha_{yx} = \frac{1}{2}\,(a_n - a_{\tau})\,\sin2\phi$,
$\alpha_{xx}-\alpha_{yy} = (a_n-a_{\tau})\,\cos2\phi$. Note the sign
alternation for $\alpha_{xy}$ and $\alpha_{xx}-\alpha_{yy}$
components.

\vspace{0.5cm}

\noindent
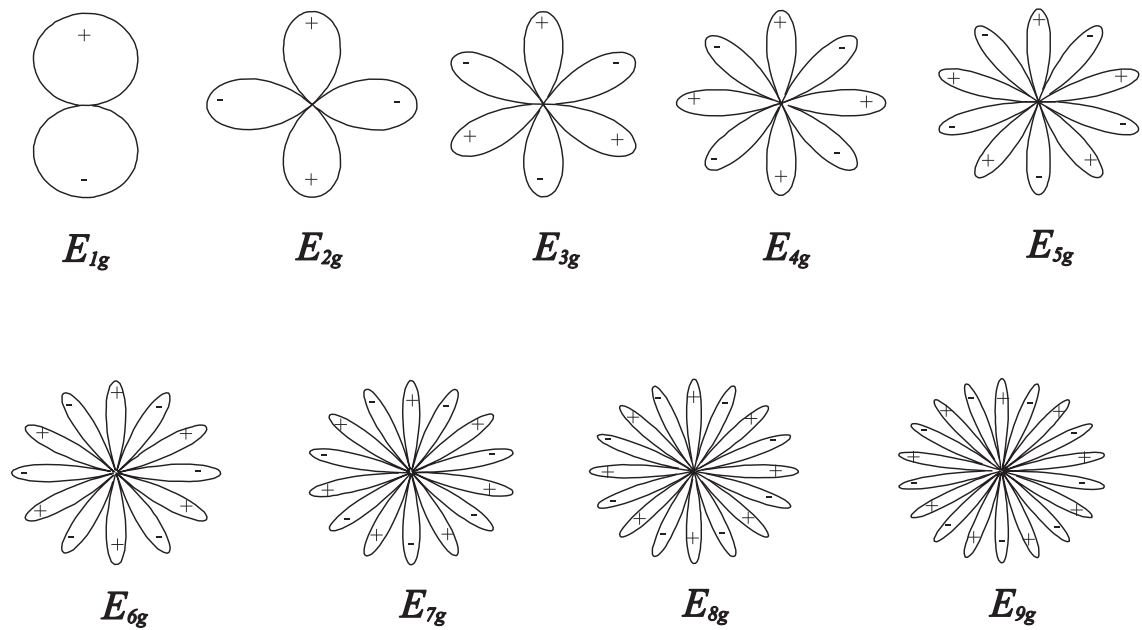
Fig.~5. Schematic representations of $E_{1g}$ -- $E_{9g}$ modes. Note
the sign alternations in lobes of vibrational modes, representing the
relative displacements of atoms from equilibrium positions.

\vspace{0.5cm}

\noindent
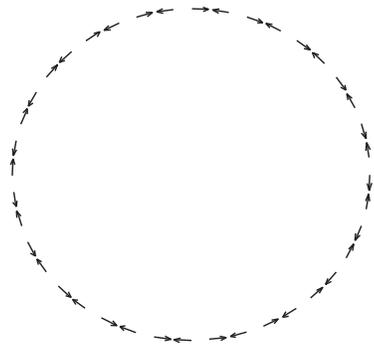
Fig.~6. Top view (in the ($x,y$)-section of $(10,10)$ CNT) of
Raman-active $A_{1g}$ modes:
(a) ``breathing'' mode ($\omega$ = 150 cm${}^{-1}$);
(b) high-frequency tangential mode ($\omega$ = 1686 cm${}^{-1}$).

\vspace{0.5cm}

\noindent
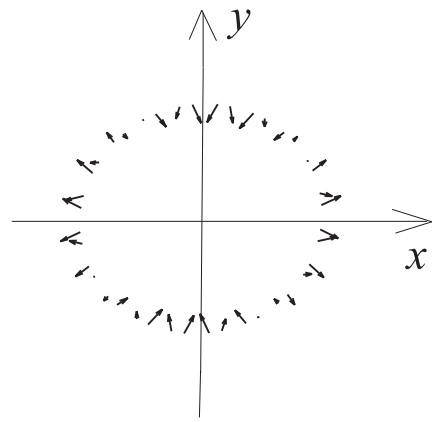
Fig.~7. All $E_{2g}$ modes for $(10,10)$ CNT in the $(x,y)$ plane:
(a) $\omega$ = 12 cm${}^{-1}$; (b) $\omega$ = 331 cm${}^{-1}$;
(c) $\omega$ = 826 cm${}^{-1}$; (b) $\omega$ = 1673 cm${}^{-1}$.

\vspace{0.5cm}

\noindent
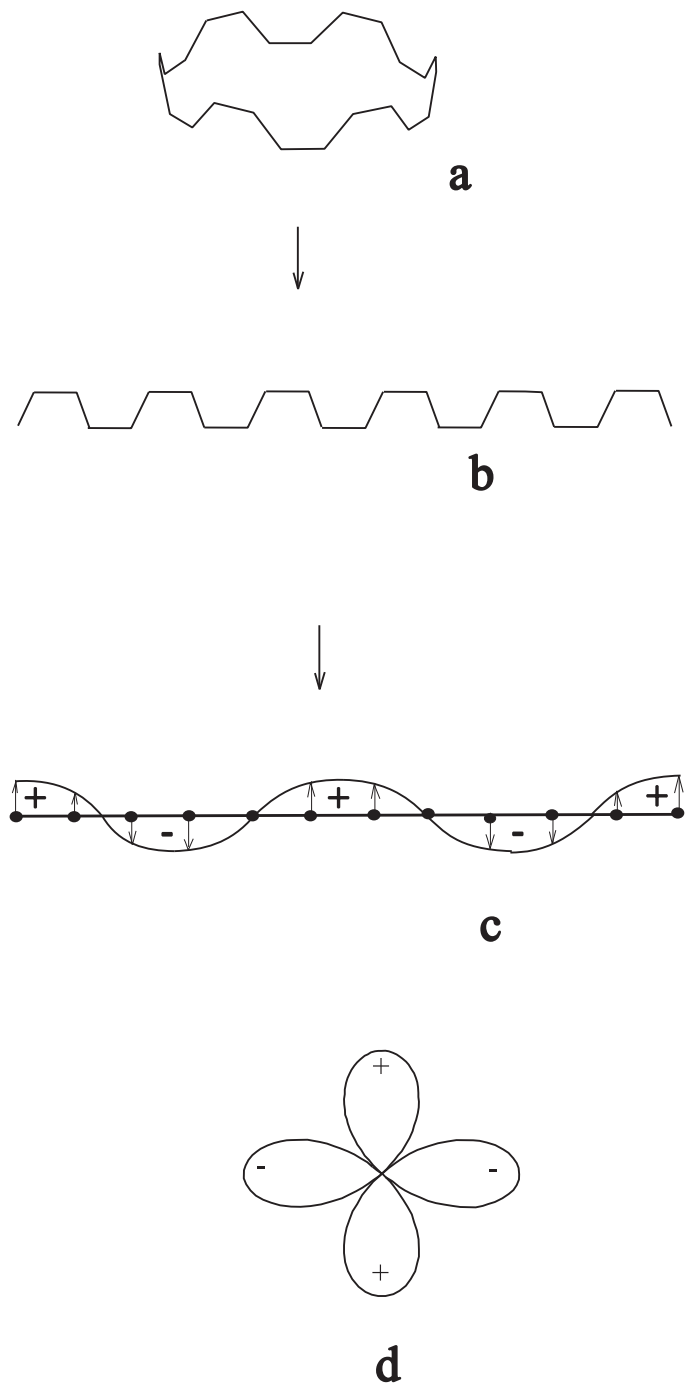
Fig.~8. (a) Armchair unit cell of $(6,6)$ CNT; (b) unit cell
projection on the plain; (c) Schematic representation of the unit
cell as a linear chain (with cyclic boundary conditions), where every
pair of atoms of one layer is substituted by one unit (the
vibrational mode for $k=2$ is shown); (d) representation of this mode
in polar coordinates.

\vspace{0.5cm}

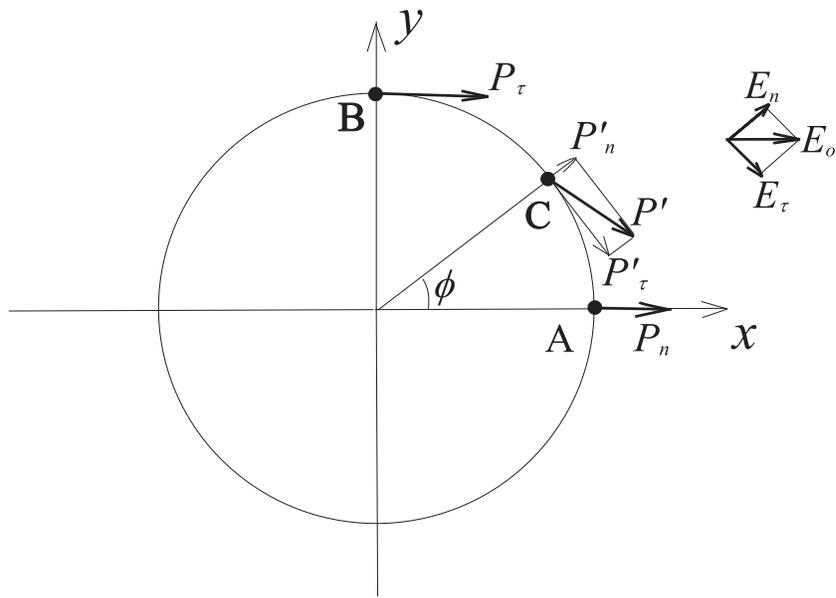
\noindent Fig.~9. Components of electric field and polarization
vectors in the cross-section of CNT for different polar angles
$\phi$.

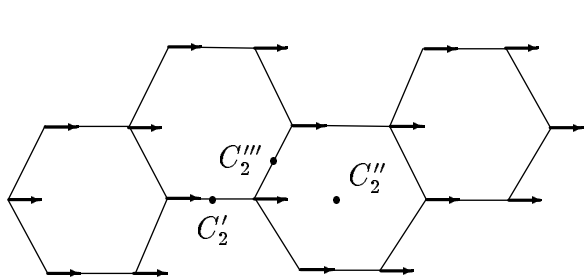
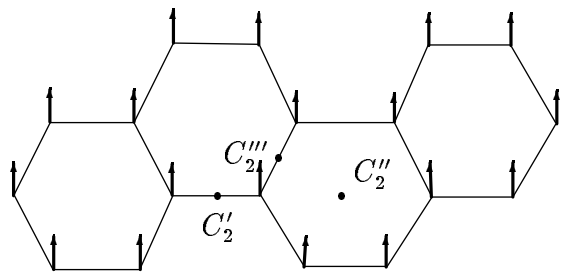
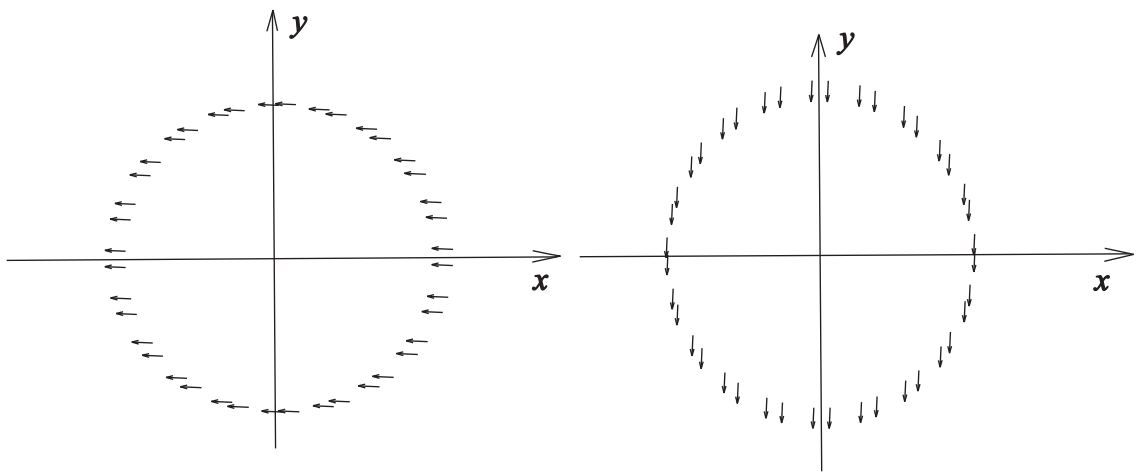
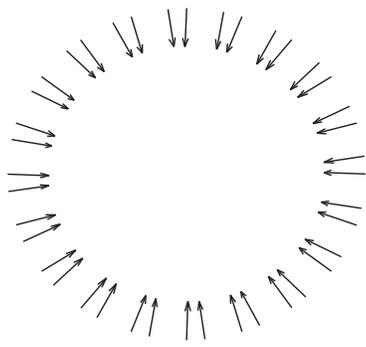
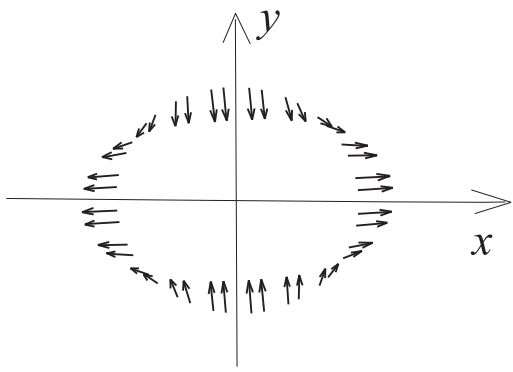
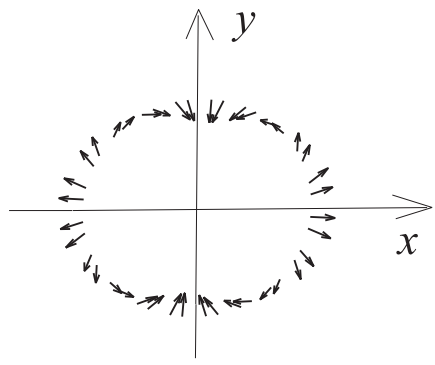
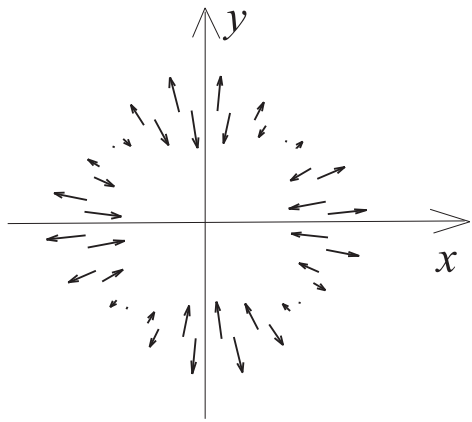
\end{document}